\documentstyle[sprocl]{article}

\input{psfig.sty}

\def\Journal#1#2#3#4{{#1} {\bf #2}, #3 (#4)}

\def\PLB{{\em Phys. Lett.}  B}
\def\PRL{\em Phys. Rev. Lett.}

\def\be{\begin{equation}}
\def\ee{\end{equation}}
\def\bea{\begin{eqnarray}}
\def\eea{\end{eqnarray}}
\newlength{\minitwocolumn}
\def\NPA{{\em Nucl. Phys.} A}
\def\PRC{{\em Phys. Rev.} C}

\begin{document}

\title{RELATIVISTIC MEAN FIELD DESCRIPTION \\
OF NUCLEAR COLLECTIVE ROTATION \\
$-$THE SUPERDEFORMED ROTATIONAL BANDS \\
IN THE A$\sim$60 MASS REGION$-$}

\author{HIDEKI MADOKORO}

\address{Department of Physics, Kyushu University, Fukuoka 812-81,
  Japan \\ E-mail: madokoro@nthp1.phys.kyushu-u.ac.jp} 

\author{MASAYUKI MATSUZAKI}

\address{Department of Physics, Fukuoka University of Education,
  Munakata, \\ Fukuoka 811-41, Japan \\ E-mail: matsuza@fukuoka-edu.ac.jp}


\maketitle\abstracts{Relativistic Mean Field Theory is applied
to the description of rotating nuclei. Since the previous formulation
of Munich group was based on a special relativistic transformation
property of the spinor fields, we reformulate in a fully covariant
manner using tetrad formalism. The numerical calculations are
performed for 3 zinc isotopes, including the newly discovered
superdeformed band in $^{62}$Zn which is the first experimental
observation in this mass region.}

In recent years, Relativistic Mean Field Theory(RMFT) has been
successful in describing various properties of nuclear matter and
ground states of finite nuclei. Some groups also tried to apply this
model to the excited states in finite nuclei. In 1989, Munich group
made a first attempt to describe rotating nuclei by combining RMFT and
the cranking assumption.~\cite{ref:KoRi89} They applied this model
mainly to the description of the superdeformed(SD) rotational bands in
the A$\sim$150, 80, and 190 mass regions.~\cite{ref:KoeRi93} In their
formulation, however, the transformation property of the spinor fields
was based on the Lorentz transformation in spite of the fact that the
rotating frame was not an inertial one. In general relativity, it is
known that the spinor fields transform as scalars under such
coordinate transformation, and hence, we think the formulation of
Munich group should be checked. Therefore, we reformulated in a fully
covariant manner using the technique of general relativity known as
tetrad formalism.~\cite{ref:We72} The resulting equations of motion
were, in fact, the same as those of Munich group. Why they obtained
the correct result was also clarified in our formulation. For detail,
see ref.\ref{ref2:MaMa97}. By solving these equations
self-consistently, we can calculate various properties of rotating
nuclei. As an application of the present model, we calculate the SD
bands in the A$\sim$60 mass region. This mass region is chosen
because, very recently, McMaster group reported the discovery of the
SD band in $^{62}$Zn, which was the first experimental observation of
the SD bands in this mass region.~\cite{ref:Sv97}

Our calculations are performed for 3 zinc isotopes, $^{60}$Zn,
$^{62}$Zn, and $^{64}$Zn using the parameter set NL-SH. Theoretically,
Ragnarsson calculated the SD states in $^{60}$Zn and predicted that
the SD minimum became yrast at $I=22$.~\cite{ref:Ra90} Experimentally,
on the other hand, the SD band in $^{62}$Zn was very recently
discovered. This band seems to become yrast at $I\geq 24$, and the
extracted $\beta_{2}$ value is 0.45. In Fig.\ \ref{fig:62Zn}(a) the
calculated moments of inertia of several SD bands in $^{62}$Zn are
shown as functions of the rotational frequency. The experimental
values are denoted by crosses. Also shown are the excitation energies
as functions of the total spin in Fig. \ref{fig:62Zn}(b). The lowest SD
band is the one named `band A'. Therefore, if we assume that this band
A corresponds to the experimentally observed one, the calculated
moments of inertia are somewhat too small compared to the experimental
values, although we can not give any decisive conclusion at the
present stage, because there are only limited number of experimental
data.

\noindent
\begin{minipage}{\minitwocolumn}
  \psfig{figure=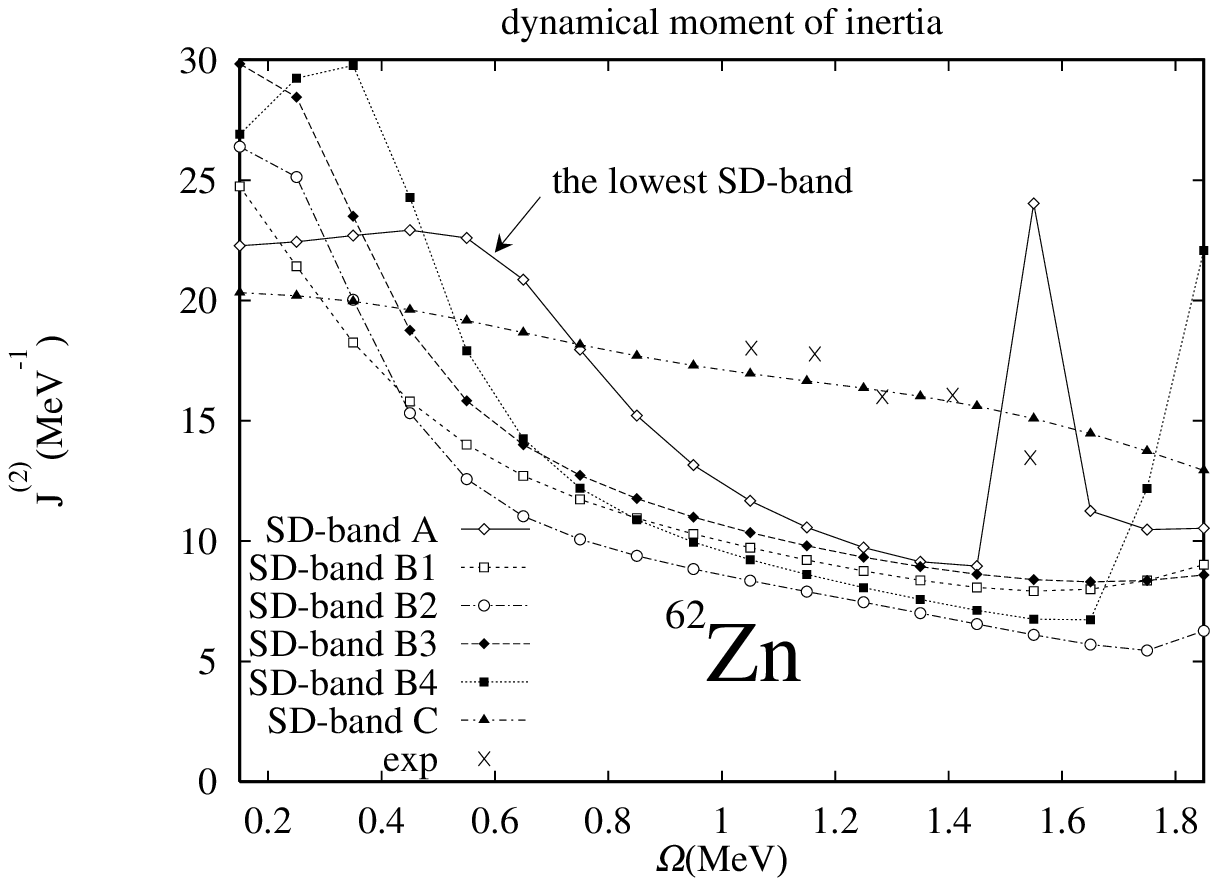,width=5.8cm,height=4.8cm}
  {\footnotesize
  Figure \ref{fig:62Zn}(a): Calculated and experimental
  dynamical moments of inertia of several SD bands in $^{62}$Zn. The
  configurations of each band are
  A:$\nu[3]^{-2}[4]^{2}\pi[3]^{-2}[4]^{2}$,
  B1$\sim$B4:$\nu[3]^{-1}[4]^{1}\pi[3]^{-2}[4]^{2}$,
  C:$\nu[3]^{-4}[4]^{4}\pi[3]^{-2}[4]^{2}$. \\
%  According to the parity and signature of the valence neutrons, band
%  B can take 4 possibilities.
  }
  \refstepcounter{figure} % The value of the counter `figure'
			  % increases by one.
  \label{fig:62Zn}
\end{minipage}
\hspace{\columnsep}
\begin{minipage}{\minitwocolumn}
  \psfig{figure=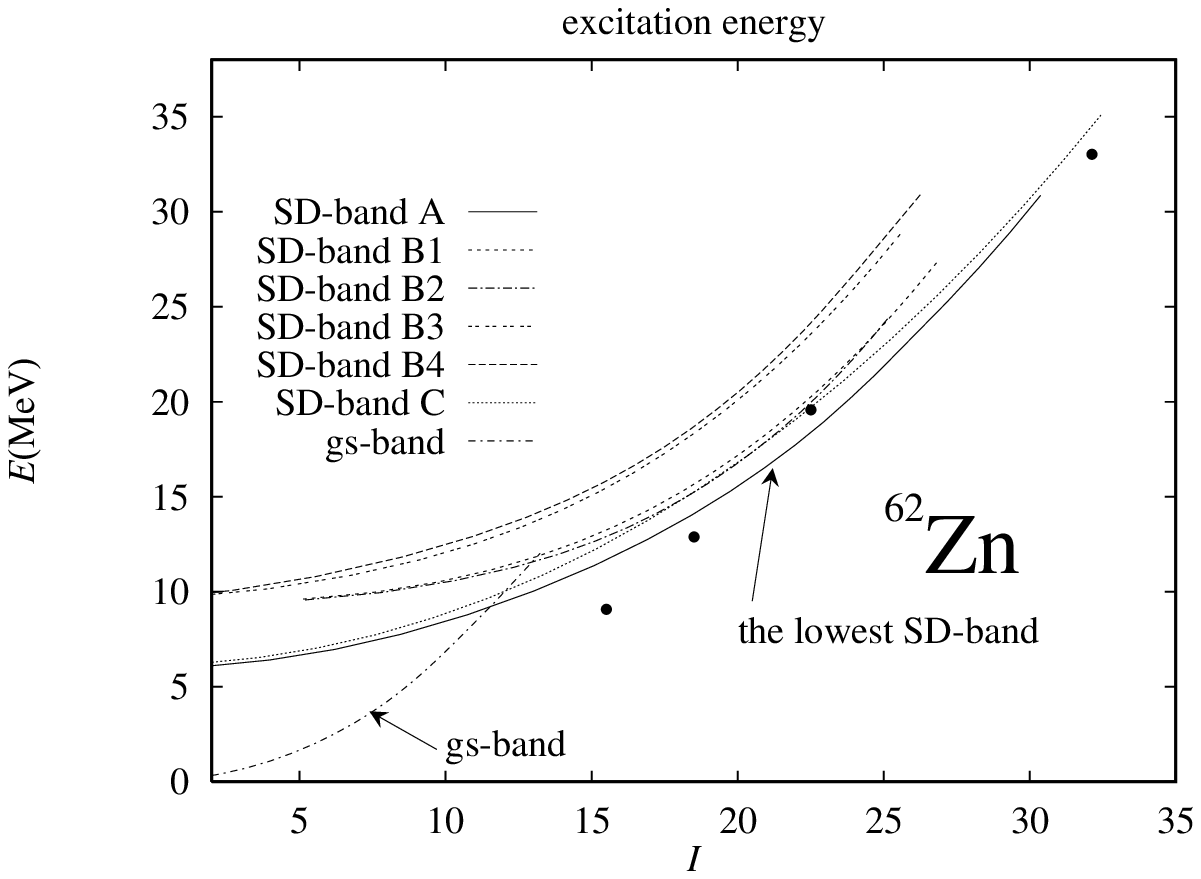,width=5.8cm,height=4.8cm}
  {\footnotesize
  Figure \ref{fig:62Zn}(b): Excitation energies of several SD bands
  together with the ground state band in $^{62}$Zn.
  The discrete points represent the oblate terminating states.
  The meaning of each SD band is the same as (a). \\
  } \\
\end{minipage}

For other isotopes, there are no experimental observations and we show
only the calculated excitation energies of the lowest one or two SD
state(s). Figure \ref{fig:other}\ shows the excitation energies as
functions of the total spin. For comparison we also show the ground
state bands and the several oblate terminating states. The lowest SD
states seem to become yrast at $I\simeq16\sim24$ in all these
isotopes. For $^{60}$Zn, this is consistent with the prediction of
Ragnarsson.

\noindent
\begin{minipage}{\minitwocolumn}
  \psfig{figure=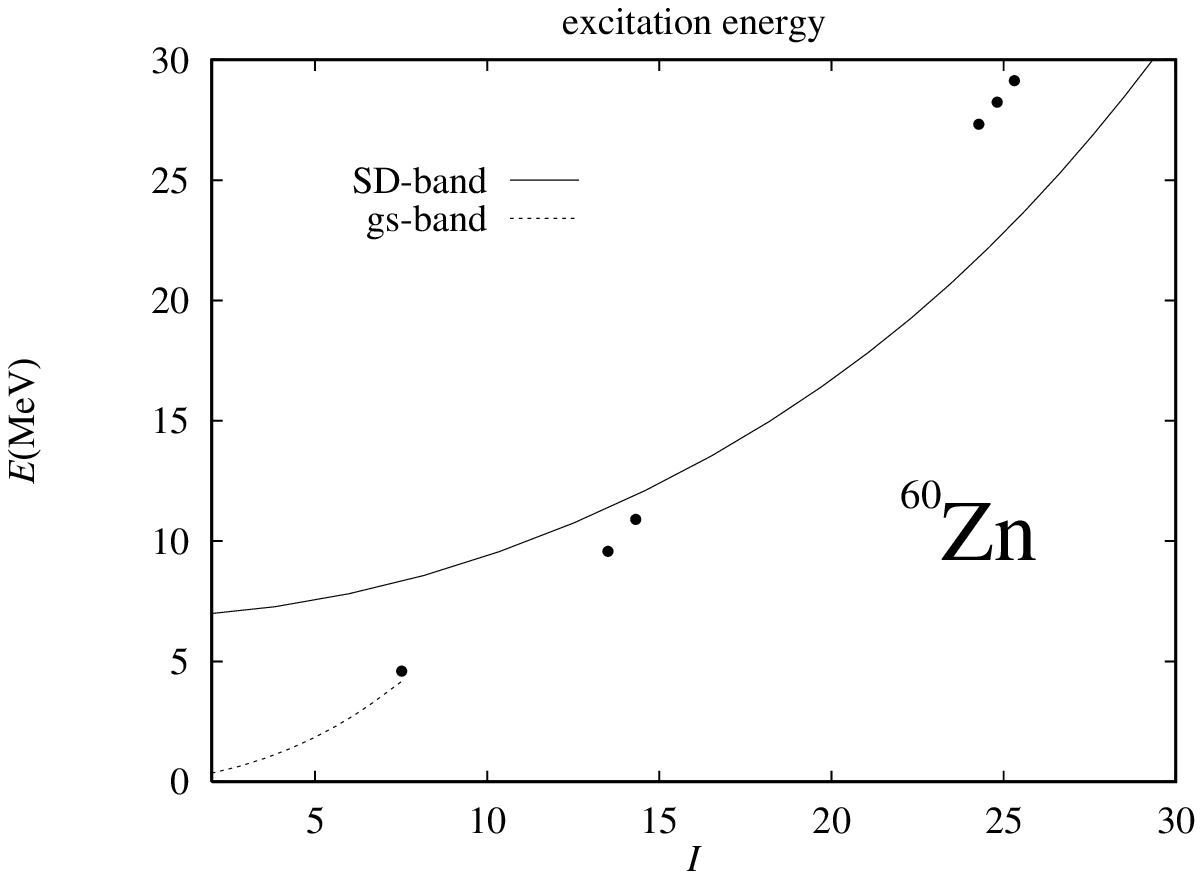,width=5.8cm,height=4.8cm}
  {\footnotesize
  Figure \ref{fig:other}(a): Excitation energies of the lowest SD band and 
  the ground state band in $^{60}$Zn. The discrete points represent
  the oblate terminating states. The configuration of the SD band is
  $\nu[3]^{-2}[4]^{2}\pi[3]^{-2}[4]^{2}$. \\
  }
  \refstepcounter{figure} % The value of the counter `figure'
			  % increases by one.
  \label{fig:other}
\end{minipage}
\hspace{\columnsep}
\begin{minipage}{\minitwocolumn}
  \psfig{figure=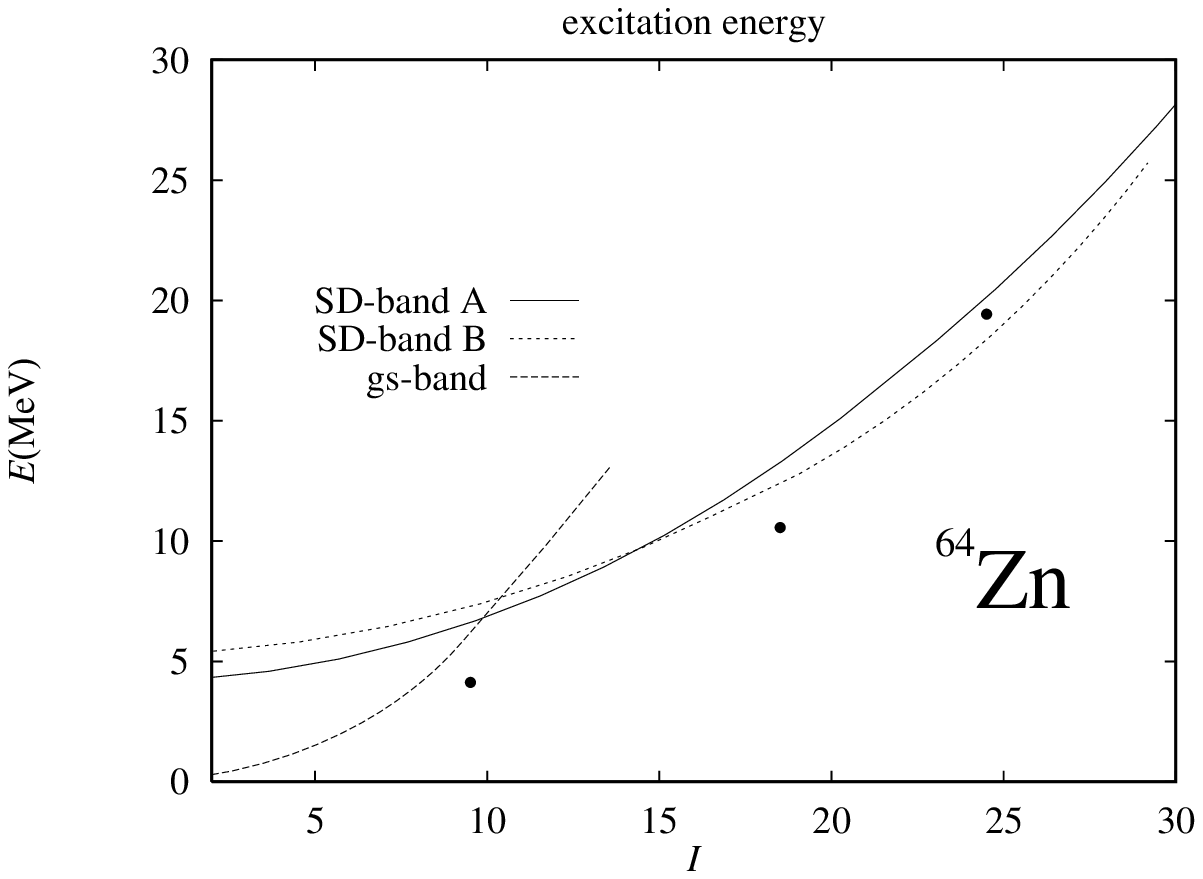,width=5.8cm,height=4.8cm}
  {\footnotesize
  Figure \ref{fig:other}(b): The same as (a) but the lowest two SD bands
  in $^{64}$Zn are shown. The configurations of the SD bands are \\
  A:$\nu[3]^{-4}[4]^{4}\pi[3]^{-2}[4]^{2}$,
  B:$\nu[3]^{-2}[4]^{2}\pi[3]^{-2}[4]^{2}$. \\
  }
\end{minipage}

To summarize, we applied Relativistic Mean Field Theory to the
description of the SD rotational bands in the A$\sim$60 mass region,
including the newly discovered SD band in $^{62}$Zn. The calculated
dynamical moments of inertia were somewhat too small compared with the
experimental values. In future, more systematic investigation both in
theoretical and experimental way will be necessary to arrive at a
definite conclusion on the SD bands in this mass region.

\end{document}